\documentstyle[12pt,openbib]{article}
\title{A non-strictly hyperbolic system for the Einstein equations with
arbitrary lapse and shift}

\author{Andrew Abrahams, Arlen Anderson, Yvonne Choquet-Bruhat\cite{YCBadd}\\ 
and James W. York, Jr.\\
         {\it Department of Physics and Astronomy}\\
         {\it University of North Carolina,  Chapel Hill 27599-3255 USA}}
\date{June 26, 1996}

\def\dzeroh{\hat\partial_0}
\def\bnabla{\bar \nabla}
\def\Boxh{\hat{\mbox{\kern-.0em\lower.3ex\hbox{$\Box$}}}}
 
\newtheorem{theorem}{Theorem}

\begin{document}
\maketitle
\vspace{-12cm}
\hfill IFP-UNC-518
 
\hfill TAR-UNC-054
 
\hfill gr-qc/9607006
\vspace{9cm}
 
\begin{abstract}
We obtain a system for the spatial metric and extrinsic curvature of
a spacelike slice that is hyperbolic non-strict in the sense of Leray
and Ohya and is equivalent to the Einstein equations.  Its characteristics
are the light cone and the normal to the slice for any choice of
lapse and shift functions, and it admits a well-posed causal Cauchy
problem in a Gevrey class of index $\alpha=2$.  The system becomes
quasidiagonal hyperbolic if we posit a certain wave equation for the lapse
function, and we can then relate the results to our previously
obtained first order symmetric hyperbolic system for general
relativity.


\end{abstract}
\newpage

\section{Introduction}
We consider, as in previous works\cite{CBY95,aacby}, the dynamics of General
Relativity as the evolution of the geometry of spacelike slices of a 
spacetime manifold $M\times R$.  We take as coframe
\begin{equation}
\theta^0=dt,\quad \theta^i=dx^i +\beta^i dt,
\end{equation}
and write the spacetime metric
\begin{equation}
ds^2=-N^2 (\theta^0)^2 + g_{ij} \theta^i \theta^j.
\end{equation}
The geometry of a slice $M_t$, $x^0=t$, is determined by its metric
${\bf \bar g}=g_{ij} dx^i dx^j$ and extrinsic curvature $K_{ij}$.  The
operator $\dzeroh$, acting on time-dependent space tensors, is defined by
\begin{equation}
\dzeroh = {\partial\over \partial t}-{\cal L}_\beta,
\end{equation}
where ${\cal L}_\beta$ is the Lie derivative with respect to the spatial
shift vector $\beta^k$.  We have
\begin{equation}
\label{d0g}
\dzeroh g_{ij}=-2N K_{ij}.
\end{equation}

By using a combination $\Omega_{ij}=\dzeroh R_{ij}- 2\bnabla_{(i}R_{j)0}$ 
of the $\dzeroh$ derivative of
$R_{ij}$ with space derivatives of the momentum constraints, we have
obtained a second-order system for $K_{ij}$. This system reduces to
a quasi-diagonal system for any 
choice of shift when the lapse $N$ is chosen so that either
the time slicing is harmonic or the slices have {\it a priori} a
given mean extrinsic curvature, $H\equiv K^k\mathstrut_k=h(x,t)$.  The system
obtained for $({\bf \bar g, K})$ is hyperbolic and causal in the first
case.  It is a mixed hyperbolic (causal) and elliptic system in
the second case, where one uses the $R_{00}$ equation to determine
$N$.  We have shown moreover that the system with the harmonic
time slicing (and also generalizations of this slicing) can be written as a 
first order symmetric hyperbolic system if one uses also the $R_{00}$
equation.

In this paper we use a combination of $\dzeroh\dzeroh R_{ij}$ and  
derivatives of the constraints $R_{i0}$ and $R_{00}$ to obtain
a system for $({\bf \bar g, K})$ whose characteristics are 
the light cone and the normal to the spatial slice, for any choice of
lapse and shift (as functions of spacetime).  This system is 
hyperbolic non-strict in the sense
of Leray-Ohya\cite{LeO}.  It admits a well-posed causal Cauchy problem in a 
Gevrey class of index $\alpha=2$.

We also show that if the lapse $N$ satisfies
a wave equation whose source is an arbitrary (smooth) function of
spacetime, ${\bf \bar g,\ K},\ N$ and first derivatives of $N$, 
the system becomes hyperbolic in the usual sense, with solutions 
in local Sobolev spaces.  This system can be put in first order
form, but it is not symmetric because the symmetrizing
matrix is semi-definite.  However, for an appropriate choice of the wave
equation for the lapse, 
one can write our new system as a symmetric hyperbolic
first order system that is the $\dzeroh$ derivative of the
one previously obtained\cite{CBY95,aacby}.  (We can also show
that if the lapse is required to satisfy an appropriate elliptic
equation, there is a well-posed hyperbolic-elliptic formulation\cite{CBY96b}.) 

\section{Non-strictly hyperbolic system}

We consider the combination
\begin{equation}
\Lambda_{ij}\equiv \dzeroh\dzeroh R_{ij} -2\dzeroh \bnabla_{(i}R_{j)0}
+\bnabla_i \bnabla_j R_{00}\equiv \dzeroh \Omega_{ij} 
+\bnabla_i \bnabla_j R_{00},
\end{equation}
where $(ij)={1\over 2}(ij +ji)$.
We use the expression for $\Omega_{ij}$, in the form given in \cite{aacby}
from the 3+1 decomposition of the Ricci tensor, together with the
expression for $R_{00}$, to obtain the following identity
\begin{eqnarray}
\label{Lam}
\Lambda_{ij}&\equiv& \dzeroh (N\Boxh K_{ij}) +\dzeroh J_{ij}+
\bnabla_i \bnabla_j(N\Boxh N-N^2 K^{mk}K_{mk})+
{\cal C}_{ij}, \nonumber
\end{eqnarray}
where
\begin{equation}
\Boxh =-N^{-1}\dzeroh N^{-1} \dzeroh  
+ \bnabla^k \bnabla_k ,
\end{equation}
\begin{eqnarray}
\label{Jij}
J_{ij}&=& \hat\partial_0 (H K_{ij} - 2 K_{i}\vphantom{|}^{k} K_{jk})
 +(N^{-2}\hat\partial_0 N+ H)\bar\nabla_i \bar\nabla_j N \nonumber \\
&&\hspace{-0.75cm}
-2N^{-1}(\bar\nabla_k N) \bar\nabla_{(i}(N K^{k}\vphantom{|}_{j)})
+3 (\bar\nabla^k N) \bar\nabla_k K_{ij} \\
&&\hspace{-0.75cm} +N^{-1}K_{ij} \bar\nabla^k (N\bar\nabla_k N)
-2 \bar\nabla_{(i}(K_{j)}\vphantom{|}^{k}\bar\nabla_k N)
+N^{-1} H \bar\nabla_i\bar\nabla_j N^2
\nonumber \\
&&\hspace{-0.75cm} 
+2 N^{-1}(\bar\nabla_{(i} H)(\bar\nabla_{j)}N^2 )
 -2N K^{k}\vphantom{|}_{(i}\bar R_{j)k}
-2N \bar R_{kijm}K^{km}, \nonumber
\end{eqnarray}
and
\begin{equation}
{\cal C}_{ij}\equiv  \bnabla_i\bnabla_j(\dzeroh (N^{-1}\dzeroh N) +
N \dzeroh H) -
\dzeroh (N^{-1} \bnabla_i\bnabla_j (\dzeroh N+N^2 H)).
\end{equation}
We see that ${\cal C}_{ij}$ contains terms of at most second order in
${\bf K}$, and first order in ${\bf \bar g}$, after replacing
$\dzeroh g_{ij}$ by $-2N K_{ij}$.

The identity given above shows that for a solution of the Einstein equations
\begin{equation}
R_{\alpha\beta}= \rho_{\alpha\beta},
\end{equation}
the extrinsic curvature ${\bf K}$ satisfies, for {\it any choice of
lapse $N$ and shift $\beta$}, a third order differential system which is
quasidiagonal with principal part the hyperbolic operator $\dzeroh \Boxh$.
The other unknown ${\bf \bar g}$ appears at second order except for
third derivatives occuring in $ \bnabla_j\bnabla_i \Boxh N$. To avoid
an explicit discussion of properties of matter evolution equations,
we consider below the vacuum case.

\begin{theorem}
Any solution of the vacuum Einstein equations $R_{\alpha\beta}=0$ 
satisfies the following system of partial differential equations
\begin{equation}
\label{sys}
\Lambda_{ij}=0, \quad \dzeroh g_{ij}=-2NK_{ij}.
\end{equation}
For any choice of shift $\beta$ and lapse $N>0$, (\ref{sys}) is a non-strictly
hyperbolic system with domain of dependence determined by the light cone 
when the metric $\bar g$ is properly Riemannian.

If the Cauchy data ${\bf \bar g}(.,0)={\bf \gamma},\ {\bf K}(.,0)={\bf k}$
belong to the Gevrey class $G^{2,{\rm loc}}_2(M_0)$, with ${\bf \gamma}$
properly Riemannian, there exists a solution of the system (\ref{sys})
in a neighborhood $U$ of $M_0$, for any $N,\beta\in G_2^{2,{\rm loc}}(U)$,
whose initial data $\dzeroh {\bf K}(.,0)$
and $\dzeroh\dzeroh{\bf K}(.,0)$ are determined by the restrictions
$R_{ij}(,.0)=0$ and $(\dzeroh R_{ij})(.,0)=0$.
 
\end{theorem}

{\it Proof.} The principal matrix is triangular with diagonal elements
either $\dzeroh$ or $\dzeroh \Boxh$.  Both of these operators are
(strictly) hyperbolic, but the system is not (quasi-diagonal) hyperbolic, 
due to the presence of non-diagonal terms in the principal matrix.  
It is, however, hyperbolic
non-strict in the sense of Leray-Ohya \cite{LeO}.  The Gevrey class
in which the system is well-posed is obtained by quasidiagonalization
\cite{CB1} and study of the multiplicity of the characteristics of the
resulting diagonal elements.  Here we obtain a quasidiagonal fourth 
order system for ${\bf \bar g}$, equivalent to the original system
for ${\bf \bar g,\ K}$, by substituting $K_{ij}= -(2N)^{-1}\dzeroh g_{ij}$ 
into  $\Lambda_{ij}$. The principal operators on the diagonal
are then $\dzeroh\dzeroh\Boxh$.  They are hyperbolic non-strict with
multiplicity two for $\dzeroh$.  The existence and uniqueness 
theorems of Leray-Ohya \cite{LeO} give the result.

\section{Equivalence to the Einstein equations}

\begin{theorem}
Let $N>0$ and $\beta$ be arbitrary in the Gevrey class $G_2^{2,{\rm loc}}$.  
Let $({\bf \bar g,\ K})$ be a solution of the system
\begin{equation}
\label{mat}
\Lambda_{ij} = \Theta_{ij} \equiv \dzeroh\dzeroh \rho_{ij}
-2\dzeroh \bnabla_{(i} \rho_{j)0} +  \bnabla_j\bnabla_i \rho_{00},
\end{equation}
where $\rho_{\alpha\beta}$ is some symmetric 2-tensor belonging
to $G_2^{2,{\rm loc}}$ that satisfies the conservation laws
\begin{equation}
\nabla_\alpha (\rho^{\alpha \beta} -{1\over 2} g^{\alpha\beta}
g^{\lambda \mu} \rho_{\lambda \mu})=0.
\end{equation}
Suppose the Cauchy data $({\bf \bar g,\ K},\ \dzeroh {\bf \bar K},
\ \dzeroh\dzeroh {\bf \bar K})_{M_0}$ satisfy on $M_0$ the Einstein
equations $(R_{\alpha\beta}-\rho_{\alpha\beta})_{M_0}=0$
together with $(\dzeroh(R_{ij}-\rho_{ij}))_{M_0}=0$.
Then the metric $({\bf \bar g}, N, \beta)$ satisfies the Einstein
equations $R_{\alpha\beta}-\rho_{\alpha\beta}=0$ in the
domain of dependence of $M_0$.
\end{theorem}

{\it Note.} The equations $(R_{ij}-\rho_{ij})_{M_0}=0$ and
$(\dzeroh(R_{ij}-\rho_{ij}))_{M_0}=0$ determine $(\dzeroh K_{ij})_{M_0}$
and $(\dzeroh\dzeroh K_{ij})_{M_0}$ when the data $({\bf \bar g, K})_{M_0}$
are given.  The equations $(R_{0\alpha}-\rho_{0\alpha})_{M_0}=0$
are then equivalent to the usual constraints on $({\bf \bar g, K})_{M_0}$,
while the conditions $(\dzeroh (R_{0\alpha}-\rho_{0\alpha}))_{M_0}=0$ are a 
consequence of the restriction to $M_0$ of the (twice-contracted)
Bianchi identities
and conservation laws.

\noindent {\it Proof.} We set 
$X_{\alpha\beta}=R_{\alpha\beta}-\rho_{\alpha\beta}$.  The Bianchi
identities, together with the conservation laws, imply
\begin{equation}
\label{bi2}
\nabla_i X^{i}\mathstrut_0 -{1\over 2}( N^{-2} \nabla_0 X_{00}
+g^{ij} \nabla_0 X_{ij})=0.
\end{equation}
We rewrite (\ref{mat})
\begin{equation}
\label{mat2}
\nabla_0\nabla_0 X_{ij} -\nabla_0 \nabla_{(i}X_{j)0} + 
\nabla_j \nabla_i X_{00}= {\rm lhe_1},
\end{equation}
where ${\rm lhe}_n$ means linear and homogeneous in ${\bf X}$ and derivatives
of order less than or equal to $n$.

The sum of twice the $\nabla_0$ derivative of (\ref{bi2}) and the
contraction with $g^{ij}$ of (\ref{mat2}) gives the equation
\begin{equation}
\label{boxX00}
-N^{-2} \nabla_0\nabla_0 X_{00} + g^{ij} \nabla_j\nabla_i X_{00}=
{\rm lhe_1}.
\end{equation}
The $\nabla_0^2$ derivative of the Bianchi identity-conservation law with 
free index $j$ reads
\begin{equation}
\label{bi3}
\nabla_0^2 \nabla_0 X^{0j} + \nabla_0^2 \nabla_i X^{ij}
-{1\over 2} g^{ij} \nabla_0^2 \nabla_i(-N^{-2} X_{00} +
g^{mk} X_{mk})=0.
\end{equation}
We use (\ref{mat2}) and the Ricci identity to obtain
\begin{eqnarray}
\label{boxXj0}
\nabla_0^2 \nabla_0 X^{0j} + 2\nabla_0 \nabla_k \nabla^{(k} 
X^{j)}\mathstrut_0 - \nabla^j \nabla^k \nabla_k X_{00} +
{1\over 2} N^{-2} \nabla^j \nabla_0^2 X_{00}\hspace{1cm}&& \\ 
-{1\over 2} \nabla^j[2\nabla_0\nabla_k X^{k}\mathstrut_0-
\nabla^k \nabla_k X_{00}] &&\nonumber \\ 
\equiv\nabla_0(-N^{-2} \nabla_0^2 X^j\mathstrut_0 + \nabla^k \nabla_k
 X^j\mathstrut_0) -{1\over 2} \nabla^j(-N^{-2} \nabla_0^2 X_{00}
+\nabla^k \nabla_k X_{00})&=&{\rm lhe_2}. \nonumber
\end{eqnarray}
The equations (\ref{mat2}), (\ref{boxX00}), (\ref{boxXj0}), with
unknowns $X_{\alpha\beta}$ have a triangular principal matrix with
diagonal principal operators $\Boxh$ and $\dzeroh \Boxh$.  It is
a non-strictly hyperbolic system.  It is equivalent to the 
quasidiagonal system consisting of (\ref{boxX00}), (\ref{bi3})
and the following equation obtained by combination of the previous
ones
\begin{equation}
\label{boxXij}
\nabla_0^2(-N^{-2} \nabla_0^2  + \nabla^k \nabla_k) X_{ij}=
{\rm lhe_3}.
\end{equation}
The system (\ref{boxX00}), (\ref{boxXj0}), (\ref{boxXij}) is hyperbolic
non-strict in the Gevrey class $G_2^{2,{\rm loc}}$.  The result follows from
the Leray-Ohya uniqueness theorem.

\section{Hyperbolic system for $({\bf \bar g, K}, N)$}

We add to the system (\ref{sys}) a wave equation for $N$ with arbitrary
right hand side.  Namely, we set
\begin{equation}
\label{boxN}
\Boxh N\equiv -N^{-1}\dzeroh (N^{-1} \dzeroh N) + \bnabla^k \bnabla_k N=F,
\end{equation}
with $F$ a given function of spacetime, $N,{\bf \bar g,K}$ and first 
derivatives of $N$.  Therefore, the
equation $\Lambda_{ij}=0$, reduced by the substitution of (\ref{boxN})
and the replacement of $\dzeroh g_{ij}$ by $-2N K_{ij}$ wherever it occurs,
takes the form
\begin{equation}
\label{d0boxK}
\dzeroh \Boxh K_{ij}=f_{ij}({\rm 2\ in\ }{\bf \bar g,K};{\rm 3\ in\ }N).
\end{equation}
The left hand side is third order in ${\bf K}$ and of order less
than 3 in ${\bf\bar g}$ and $N$. The numbers in $f_{ij}$ denote the
order of the highest derivatives occuring there. 
Because $\Boxh N$ involves first derivatives of ${\bf \bar g}$, to
obtain a hyperbolic system we differentiate (\ref{boxN})
by $\dzeroh$ and use  $\dzeroh g_{ij}=-2N K_{ij}$. 
This gives an equation which is third order
in $N$ and first order in ${\bf \bar g}$ and ${\bf K}$,
\begin{equation}
\label{d0boxN}
\dzeroh \Boxh N= \dzeroh F.
\end{equation} 

\begin{theorem}
The system (\ref{d0g}), (\ref{d0boxK}), (\ref{d0boxN}), 
is a quasi-diagonal hyperbolic system.  
\end{theorem}

We attribute to the unknowns and equations the following Leray-Volevic
indices:
\begin{eqnarray}
m({\bf K})=3,\quad m({\bf \bar g})=3, \quad m(N)=4, \\
n(\ref{d0boxK})=0,\quad n(\ref{d0g})=2,\quad n(\ref{d0boxN})=1. \nonumber
\end{eqnarray} 
The system assumes then a quasidiagonal form\cite{prin} with principal 
operators $\dzeroh$ and $\dzeroh \Boxh$.  It is a hyperbolic system.

We remark that a mixed hyperbolic-elliptic system for $({\bf \bar g,
K},N)$, with $N$ determined by an elliptic equation is also possible
\cite{CBY96b}.
It is a straightforward extension of the result in \cite{CBY96a}.

\section{First order  system}

The system (\ref{d0g}), (\ref{boxN}), (\ref{d0boxK}) can be
put in first order form by a straightforward procedure.  The
system can be symmetrized, but the symmetrizing matrix is
found to be semi-definite, not positive definite, so the
usual existence and uniqueness theorems do not apply.  It is
an open question whether the system in first-order form is
well-posed.

To achieve a symmetric hyperbolic system, the source 
$F$ of the wave equation (\ref{boxN}) for $N$ can be chosen so
that no second spatial derivatives of ${\bf K}$ appear in the 
first order system:
\begin{equation}
\label{boxNsymm}
\Boxh N -NK_{mk}K^{mk} +N H^2=\tilde F({\bf \bar g},N,b_0,a_i;x,t).
\end{equation}
This equation is closely related to a form of harmonic time slicing that
is allowed in our second order system for ${\bf K}$ \cite{CBY95,aacby},
namely,
\begin{equation}
\label{d0N}
\dzeroh N + N^2 H=Nf(\sqrt{{\bf \bar g}}\alpha^{-1},N;x,t),
\end{equation}
where $\alpha(x,t)$ is an arbitrary positive scalar density of weight
one and $f$ is a scalar.  Using (\ref{d0N}) and the Einstein equation
for $R_{00}$ gives (\ref{boxNsymm}) with $\tilde F=\tilde F(
\sqrt{{\bf \bar g}}\alpha^{-1},N,b_0;x,t)$, where $b_0=N^{-1}\dzeroh N$ 
is related to $H$ by
the harmonic condition (\ref{d0N}).   In this case, the symmetric system
obtained using (\ref{boxNsymm}) is the time derivative of the
first order symmetric hyperbolic system previously obtained in
\cite{CBY95,aacby}.

Acknowledgments.  A.A., A.A., and J.W.Y. were supported by National
Science Foundation grants PHY-9413207 and PHY-9318152/ASC-9318152
(ARPA supplemented).


\begin{thebibliography}{99}
\bibitem{prin}  The principal operator for the $p^{\rm th}$ unknown
in the $q^{\rm th}$ equation is of order $m(p)-n(q)$.
\end{thebibliography}
\end{document}